\documentclass[preprint,showpacs,showkeys]{elsarticle}
\usepackage[T1]{fontenc}
\usepackage{bm}
\usepackage{graphicx}
\usepackage{amssymb}
\usepackage{amsfonts}
\usepackage{amsmath}
\usepackage{multirow}
\usepackage{pstricks}
\usepackage{graphicx}
\begin{document}

\title{MBE grown microcavities based on selenium and tellurium compounds}

\author{J.-G.~Rousset}

\author{T.~Jakubczyk}

\author{J.~Kobak}

\author{R.~Rudniewski}

\author{P.~Piotrowski}

\author{M.~\'{S}ciesiek}

\author{E.~Janik}

\author{J.~Borysiuk}

\author{T.~Slupinski}

\author{A.~Golnik}

\author{P.~Kossacki}

\author{M.~Nawrocki}

\author{W.~Pacuski}

\address{Institute of Experimental Physics, Faculty of Physics, University of Warsaw, ul. Ho\.za 69, PL-00-681 Warszawa, Poland}



\begin{abstract}
In this work, we present three groups of microcavities: based on selenium compounds only, based on tellurium compounds only, and structures based on mixed selenium and tellurium compounds. We focus on their possible applications in the field of optoelectronic devices and fundamental physics (VCSELs, narrow range light sources, studies of cavity-polariton electrodynamics) in a range of wavelength from 540 to 760 nm.
\end{abstract}

\begin{keyword}
A3. Molecular beam epitaxy \sep B1. Cadmium compounds  \sep B1. Zinc compounds  \sep B1. Tellurides  \sep B2. Semiconducting II-VI materials  \sep B2. Magneto-optic materials
\PACS 78.20.Ci \sep 78.67.Pt \sep 68.37.Hk \sep 78.55.Et

\end{keyword}

\maketitle


\section{Introduction}

Light-matter coupling in semiconductor microcavities has been intensively investigated over the past twenty years \cite{Weisbuch_92,Gerard_1998,LeSiDang_1998}. The optimization of the growth processes and the design of microcavities containing quantum dots (QDs) or quantum wells (QWs), led to the achievement of various devices of great interest for optoelectronics applications such as vertical cavity surface emitting lasers (VCSELs) \cite{Murtagh_2004,Kruse_2004} or resonant cavity light emitting diodes (RCLEDs) \cite{Vilokkinen_2000}. From the point of view of fundamental physics of cavity electrodynamics, etched pillar microcavities containing QDs allow for controlling spatial, energetic and temporal properties of light emission \cite{Gerard_1998,Gerard_1999,Wiersig_2009,Dousse_2010,Jakubczyk_2012}. In such structures, the coherent coupling between light and matter is expected to lead to the achievement of devices of particular interest for quantum information processing \cite{Dousse_2010}. The majority of the most remarkable results in semiconductor photonics and the most important applications in the optoelectronics industry have been achieved for structures based on III-V compounds, for example GaAs \cite{Deng_science_2002,Amo_nature_2009,Dousse_2010}. However, compared to III-V compounds, the semiconductors of group II-VI exhibit stronger carrier confinement and more robust excitonic states which  is expected to extend the functionalization of the devices and the fundamental investigations on light matter coupling to higher temperatures  \cite{Kruse_2008}. In addition, II-VI compounds based emitters are good candidates to answer the problem of the low efficiency of III-V based emitters in the green-yellow range \cite{Krames_2007,Mueller2009}.

\begin{figure}[!h]
  \includegraphics[width=1\linewidth]{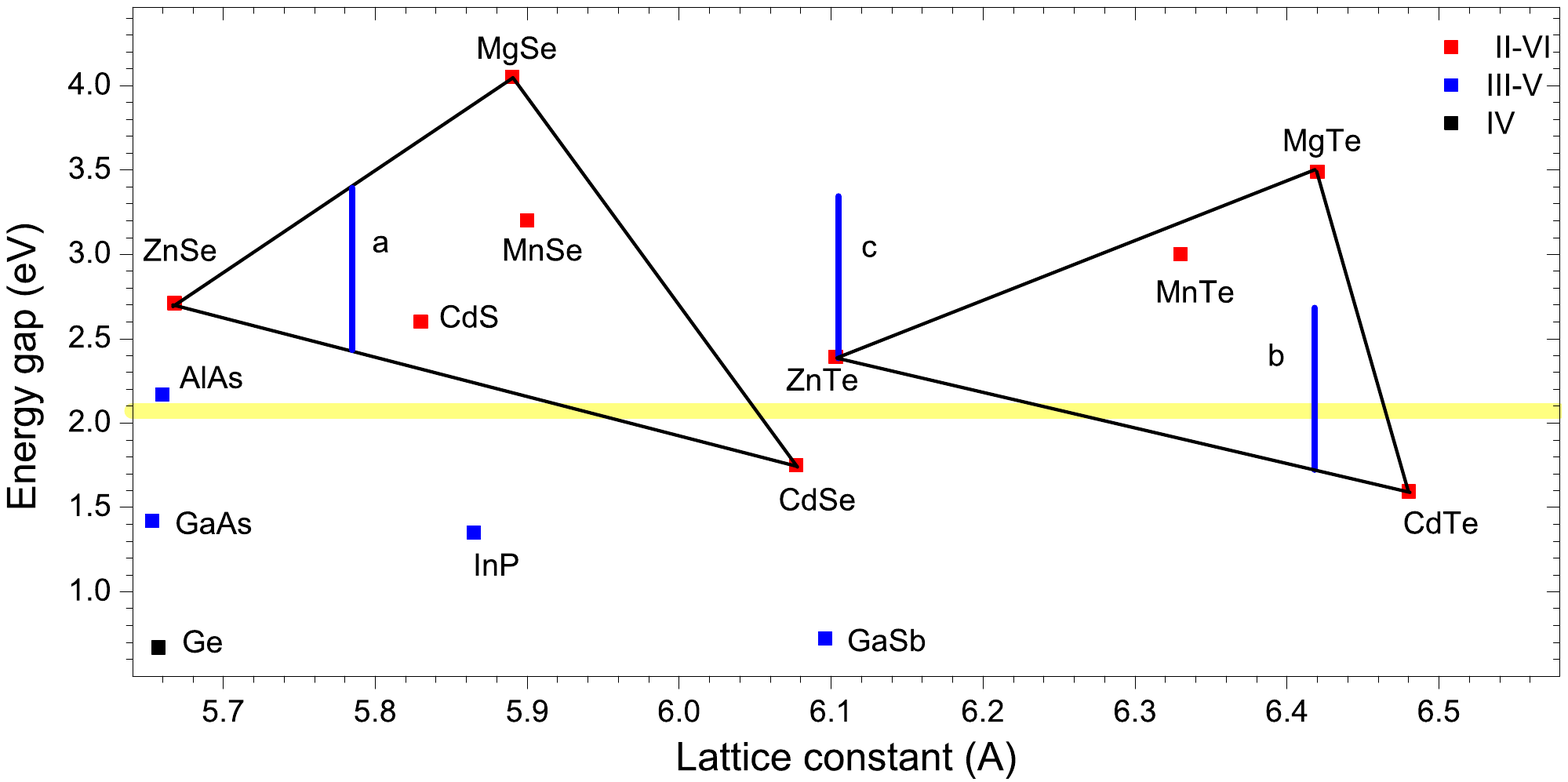}\\
\centering
\caption{(Color online) Graph showing the energy gap as a function of the lattice constant for various II-VI compounds focusing on selenium and tellurium compounds. Other materials are presented in order to point out commercially available substrates. The horizontal yellow stripe figures the energy range of interest for the yellow optoelectronics (about 2.1 eV). The vertical lines figure the structures grown with lattice matched thin layers for the three material systems discussed in the present work: a) structures based on selenides only, b) structures based on tellurides only, c) structures based on mixed selenium and tellurium compounds.}
\label{eg_lat_cst}
\end{figure}

In this work, we present the design, characterization and comparison of planar and pillar microcavities based on II-VI compounds for three material systems, focusing on the possible applications for optoelectronic devices, fundamental studies of cavity electrodynamics and materials for spintronics.  The three material systems in which we have designed and grown microcavities are schematically presented in Fig.~\ref{eg_lat_cst}. Each system is lattice matched to a different lattice parameter (Zn,Mg)Se, ZnTe, or MgTe. In structures based on selenium compounds only (Fig.~\ref{eg_lat_cst}-a), the distributed Bragg reflectors (DBRs) consist of a stack of (Zn,Cd)Se and (Zn,Mg)Se thin layers, and in the ZnSe cavity CdSe QDs are embedded. Structures based on tellurium compounds only (Fig.~\ref{eg_lat_cst}-b) are made out of (Cd,Zn,Mg)Te thin layers with various Mg contents, allowing for the tuning of the energy gap and refractive index of the layers letting the lattice constant unchanged, and with unstrained (Cd,Zn)Te QWs embedded in the cavity \cite{Rousset_2013}. In the case of structures based on mixed selenium and tellurium compounds (Fig.~\ref{eg_lat_cst}-c), the DBRs consist of ZnTe and a short period ZnTe|MgSe|ZnTe|MgTe superlattice resulting in a four compound digital alloy \cite{Pacuski_2009}; CdTe QDs are embedded in the ZnTe cavity.

\section{Design and growth}

The DBRs, microcavities, QDs and QWs were grown by molecular beam epitaxy (MBE) technique. The MBE machine provided by SVT Associates consists of two growth chambers (for III-V and II-VI semiconductor compounds), each equipped with a reflectivity setup for in situ measurements. Monitoring the reflectivity spectra during the growth allows us to verify the growth rate and thickness of the layers, which is a crucial parameter for the growth of microcavities or DBRs \cite{Kruse_2002, Shenk_2002,Mizutani_2006,Biermann_2011}. The structures are grown on GaAs:Zn (100) oriented substrates.

The design and epitaxial growth of DBRs and microcavities relies on the compromise between two opposite constrains: (i) a high refractive index contrast between the alternating layers of the distributed Bragg reflector (DBR) which is directly linked to the width of the photonic stopband and which is important for the quality factor of the cavity, and (ii) the lattice matching to avoid relaxation in thin layers leading to the formation of defects affecting the optical properties of the structure and the lifetime of QDs / QWs excitons \cite{Yeh_1988,Vahala_2004}. For a microcavity containing emitters, an additional constrain is the tuning of the emitters emission to the optical mode of the cavity. For each material system, we present the emitters that can be embedded in the cavity. During the growth of the DBRs and cavities, the substrate is not rotated. This results in a gradient of the thickness of the layers and following, an in-plane spatial distribution of the photonic stopband and cavity mode on the sample.

\subsection{selenium material system}

\begin{figure}
  \includegraphics[width=0.7\linewidth]{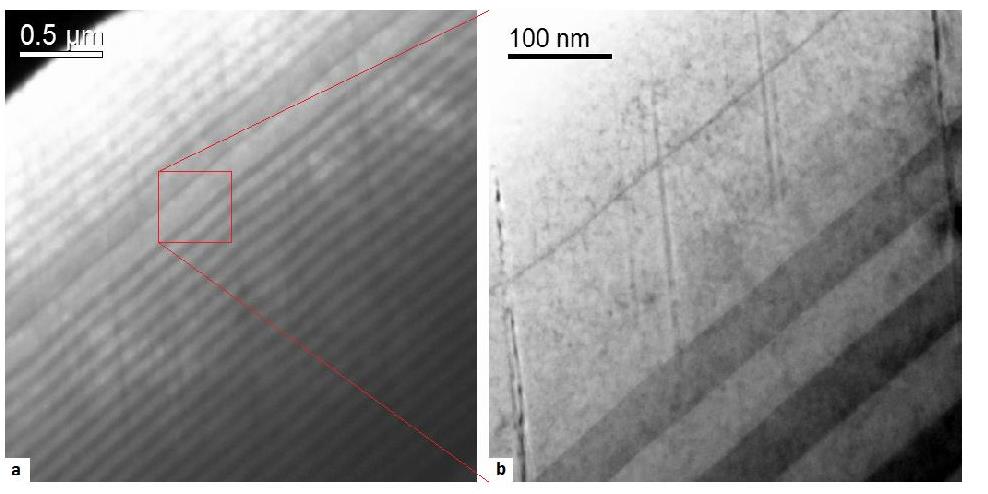}\\
\centering
\caption{Transmission electron microscopy imaging of a Se based microcavity. The steep interfaces indicate the good quality of the structure and the lattice matching of the layers in the DBR. However, the lattice mismatch between DBRs and the ZnSe cavity results in a worse crystalline quality with the formation of defects which disturbs the light emission from QDs embedded in the cavity.}
\label{TEM}
\end{figure}

The DBR consists of a 17 period stack of Zn$_{0.74}$Cd$_{0.26}$Se for the high refractive index and Zn$_{0.55}$Mg$_{0.45}$Se for the low refractive index layers with respective thickness $e_{high}=51.8$~nm and $e_{low}=55.9$~nm (see Fig.~\ref{eg_lat_cst} a), with the stopband centered at $\lambda_0=570$~nm. About 270~nm thick ZnSe buffer grown on the GaAs substrate ensures a good crystalline quality of the structure (Fig.~\ref{TEM}) revealed by transmission electron microscopy (TEM). The advantage of this structure is the low lattice mismatch with ZnSe ($\Delta a/a \approx 2\%$) which, we thought, could allow the growth of a ZnSe cavity with CdSe/ZnSe QDs \cite{Kobak_ArXiv_2013}. We did not attempt to grow a fully lattice matched structure to Zn$_{0.74}$Cd$_{0.26}$Se since it requires an other new technology - CdSe/Zn$_{0.74}$Cd$_{0.26}$Se QDs. For the structures based on selenides, the growth was realized at a substrate temperature of $T_g=335^{\circ} C$. As seen in Fig.~\ref{TEM}, the TEM imaging of such a Se based microcavity shows steep interfaces between the layers of the DBR. As shown on the inset, the lattice mismatch with ZnSe, however small, results in a worse crystalline quality and the formation of stacking faults which prevents from an efficient light emission.

As seen in Fig.~\ref{DBRs}-a, the refractive index contrast determined from the fit is $\Delta n \approx 0.20$ which might be insufficient to obtain a very high reflectivity (over $95\%$) and following, a  high quality factor for a cavity based on such DBRs. However, the position of the absorption edge at about $500$~nm, allows for applications in a wide range of the visible spectrum: green, yellow, red, infrared.

For this material system, we have grown CdSe/ZnSe QDs emitting in the range from $480$ nm to $580$ nm, with a peak position in the range of $500$~nm to $550$~nm (see Fig.~\ref{emit} a).

Comparing to selenium based DBR systems reported in the literature, our system (i) is designed for another lattice parameter - Zn$_{0.55}$Mg$_{0.45}$Se instead of GaAs \cite{Kruse_2008} or InP \cite{Peiris_1999}, (ii) it requires only ternary compounds, avoiding reported quaternary compounds \cite{Kruse_2008, Peiris_1999}, (iii) it allows to avoid sulfur \cite{Kruse_2008}, (iv) it has been applied not only for the growth of DBRs \cite{Peiris_1999} but also for the realization of entire microcavities.

    \subsection{Mixed selenium and tellurium material system}

The DBR consists of a 30 period stack of ZnTe for the high refractive index and the low refractive index digital alloy obtained by growing a short period superlattice ZnTe|MgSe|ZnTe|MgSe layers with respective thickness $e_{high}=47.2$~nm and $e_{low}=55.2$~nm  (see Fig.~\ref{eg_lat_cst} b), with the stopband centered at $\lambda_0=585$~nm. About 400 nm thick ZnTe buffer grown on the (100) GaAs substrate ensures a reasonable crystalline quality of the structure. The refractive index contrast determined from the fit is $\Delta n \approx 0.45$, allowing to reach a higher reflectivity (close to $95\%$) despite the proximity of the absorption edge at about $530$ nm. Before this work, ZnTe based DBRs \cite{Pacuski_2009} were tested only for the red - orange spectral range (590 nm $<\lambda_0<660$ nm), here we show the possibility to obtain the cavity mode for wavelength as short as $\lambda_0 \approx 565$ nm (see Fig.~\ref{MCs} b).

For this material system, we have grown CdTe/ZnTe QDs exhibiting PL emission from $560$ to $620$ nm (see Fig. \ref{emit} b). These nanostructures are particularly interesting for applications in the yellow range optoelectronics ($\lambda=575-595$~nm), and also for fundamental studies of nanostructures exhibiting magneto-optical features\cite{Kobak_ArXiv_2013,Kobak_APPA_2009,Gietka_APPA_2012,Papaj_APPA_2012,Kobak_JCG_2013}. Since CdSe is also lattice matched to ZnTe, we also have grown a CdSe/(Cd,Mg)Se QW emitting at $\lambda \approx 720$~nm. In this range, the low absorption of the DBR should allow to reach a high quality factor. In addition, the high intensity of the CdSe QW emission gives hope to obtain strong photon-exciton coupling.

The temperature of the substrate was set at $366^{\circ} C$ for the DBRs and $356^{\circ} C$ for the CdTe QDs.

    \subsection{Tellurium material system}

\begin{figure}[!h]
  \includegraphics[width=0.7\linewidth]{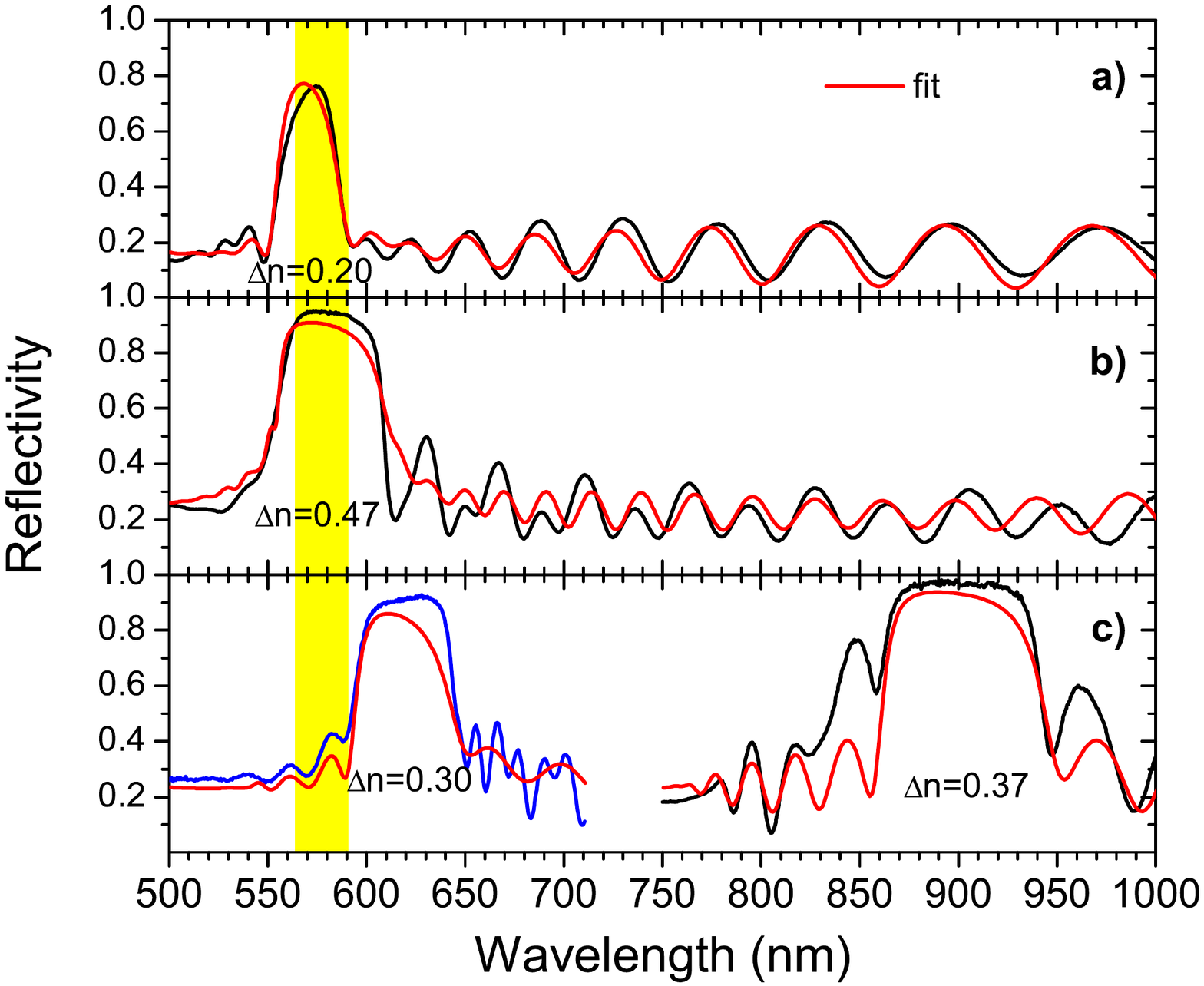}\\
\centering
\caption{(Color online) Reflectivity spectra and fit using the transfer matrix method (TMM) of DBRs for the three material systems considered. The vertical yellow stripe figures the yellow range of the visible spectrum. The refractive index contrasts $\Delta$~n are obtained from the TMM fit.}
\label{DBRs}
\end{figure}

The DBR consists of a 20 period stack of (Zn,Cd,Mg)Te layers with different Mg concentrations for the high and low refractive index layer. The different Mg concentrations are obtained thanks to two Mg sources giving different equivalent beam pressures and resulting in Mg concentrations of $\approx 10\%$ for the high refractive index layer and $\approx 50\%$ for the low refractive index layer. The whole structure is lattice matched to MgTe so that the energy gap and refractive index of the layers can be tuned through the Mg content leaving the lattice constant unchanged \cite{Rousset_2013}. As shown in Fig. \ref{DBRs} c, the range of application of such DBRs starts from $\lambda \approx 630$ nm, in this case the refractive index contrast and maximum reflectivity are lower (higher Mg concentration in the high refractive index layer).

\begin{figure}[!h]
  \includegraphics[width=0.7\linewidth]{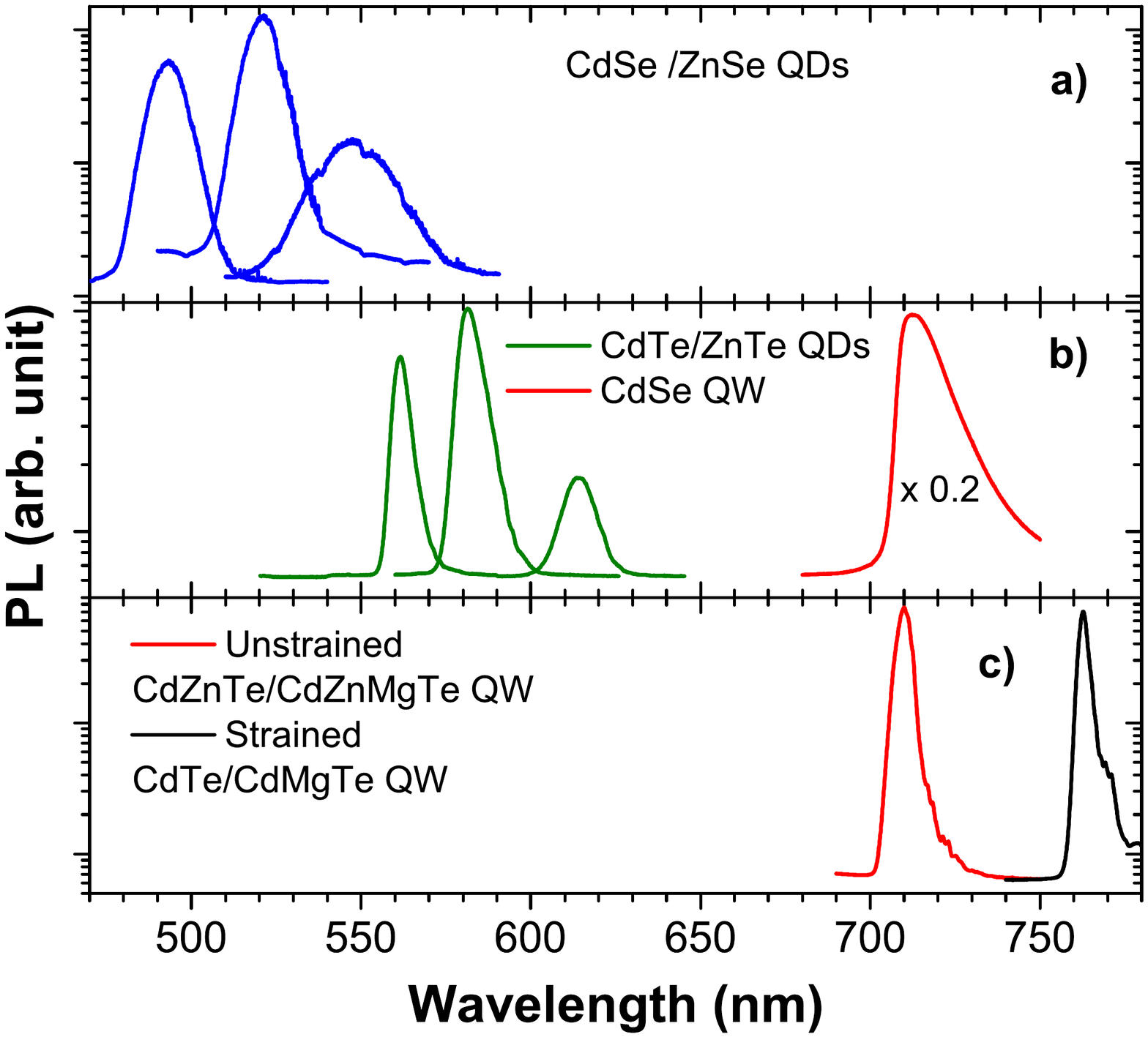}\\
\centering
\caption{(Color online) Photoluminescence spectra of emitters for each material considered. a) CdSe/ZnSe QDs for the selenium compounds only system, b) CdTe/ZnTe QDs and CdSe/ZnTe QW for the mixed selenium and tellurium compounds system, c) strained CdTe or unstrained (Cd,Zn)Te QW for the tellurium compounds only system}
\label{emit}
\end{figure}

Concerning the emitters for this material system, we face the following dilemma: one can grow a two compounds, but strained CdTe/(Cd,Zn)Te QW (or CdTe/(Cd,Zn,Mg)Te QW) or a three compounds but unstrained (Cd,Zn)Te/(Cd,Zn,Mg)Te QW \cite{Kossacki04,Rousset_2013}. We chose the three compounds QW in order to have a monolithic structure. This also gives the opportunity to grow a high number of identical QWs since no relaxation due to the lattice mismatch should occur. As seen in Fig.~\ref{emit} c, one can tune the emission energy of the (Cd,Zn)Te QW through the Zn content. The comparison of a strained CdTe/(Cd,Mg)Te and unstrained (Cd,Zn)Te/(Cd,Zn,Mg)Te QW shows that both the intensity and linewidth of the PL intensity are similar. Compared to other Te-based structures reported in the literature, our is designed without Mn in the DBR layers \cite{Kasprzak_2006} and the (Cd,Zn)Te QW is also lattice matched to the cavity \cite{Rousset_2013}. In addition, we show the possibility to obtain a photonic stopband for shorter wavelengths (see Fig.~\ref{DBRs} c); and the use of a second Mg source accelerates the growth process.

All the structures based on tellurides were grown at a temperature of $346^{\circ} C$.

\section{Microcavities}

\begin{figure}[!h]
  \includegraphics[width=0.7\linewidth]{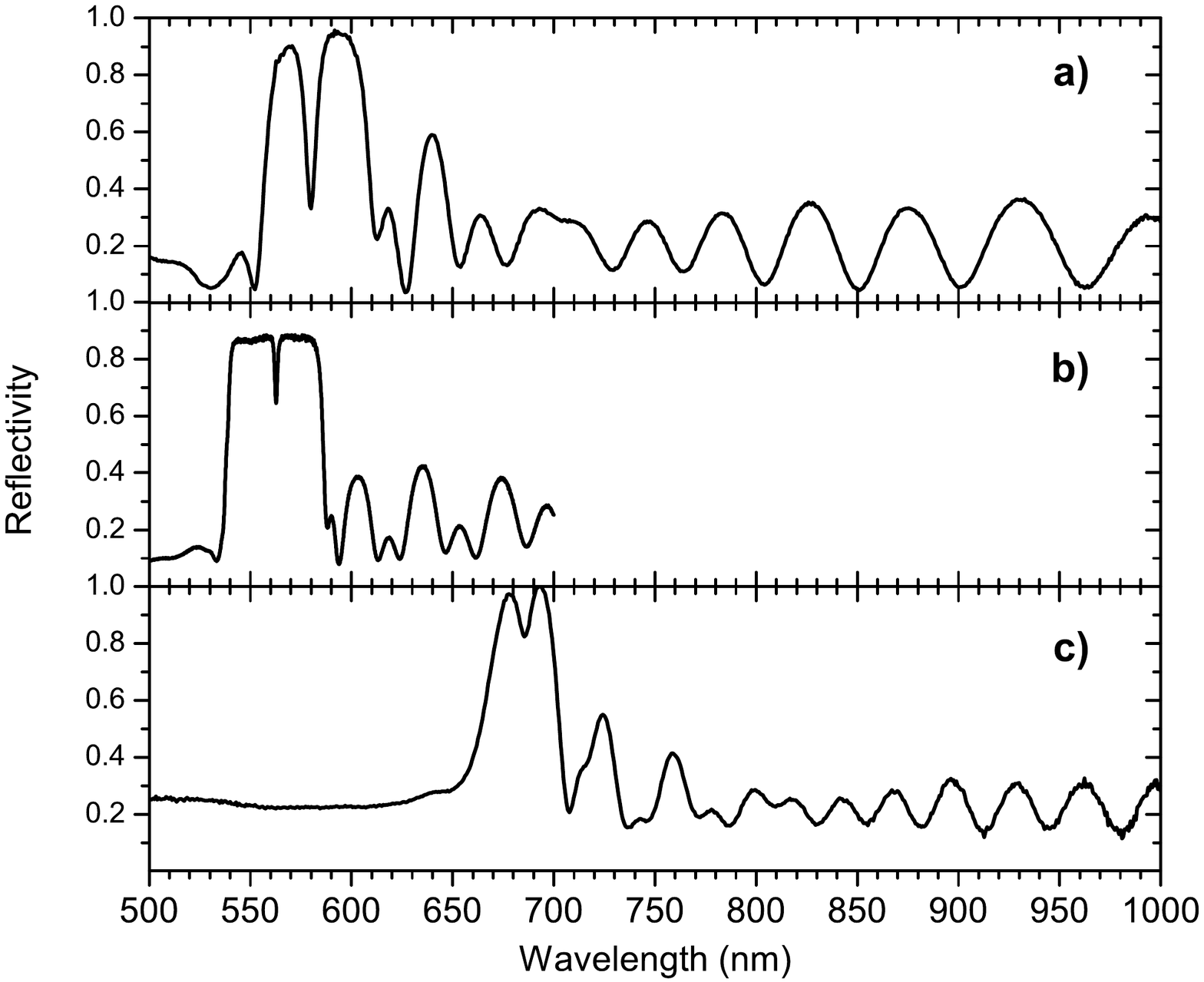}\\
\centering
\caption{Reflectivity spectra of chosen microcavities for the three material systems: a) in the Se based system the low refractive index results in a low quality factor ($Q \approx 100$), b) in the mixed Se/Te system Q is higher and the cavity can be observed for wavelengths as short as $563$ nm, c) microcavities based on Te compounds are well suited for embedding magnetoopticaly active QWs.}
\label{MCs}
\end{figure}

For each material system, we have grown and characterized microcavities by reflectivity and photoluminescence measurements. As shown in Fig.~\ref{MCs} microcavities based on selenium compounds exhibit a low quality factor resulting from the low refractive index contrast. Cavities based on tellurium compounds exhibit a similar quality factor but could be an interesting tool for the magneto-optical study of semimagnetic nanostructures such as (Cd,Mn)Te QWs embedded in a cavity since the DBRs are free of Mn (strong magneto-optical features) in contrary to the structure described in Ref.~\cite{Kasprzak_2006}. In addition, since the whole structure is lattice matched, it is theoretically possible to embed as many QWs as we want in the cavity, which gives hope to reach the ultra strong photon-exciton coupling regime \cite{Anappara_PRB_2009}. The best quality factor was obtained for microcavities in the mixed selenium/tellurium system. Such structures have already been studied \cite{Jakubczyk_APPA_2009, Kruse_2011, Jakubczyk_CEJP2011,Sciesiek_APPA_2011, Jakubczyk_JAP_2013}, but we extend the range of application to shorter wavelengths (down to $563$~nm - Fig.~\ref{MCs} b). We will then focus on the results concerning planar and pillar cavities based on mixed selenium and tellurium compounds since combine a wide wavelength application range and a high refractive index contrast.

    \subsection{planar microcavities}

\begin{figure}[!h]
  \includegraphics[width=0.7\linewidth]{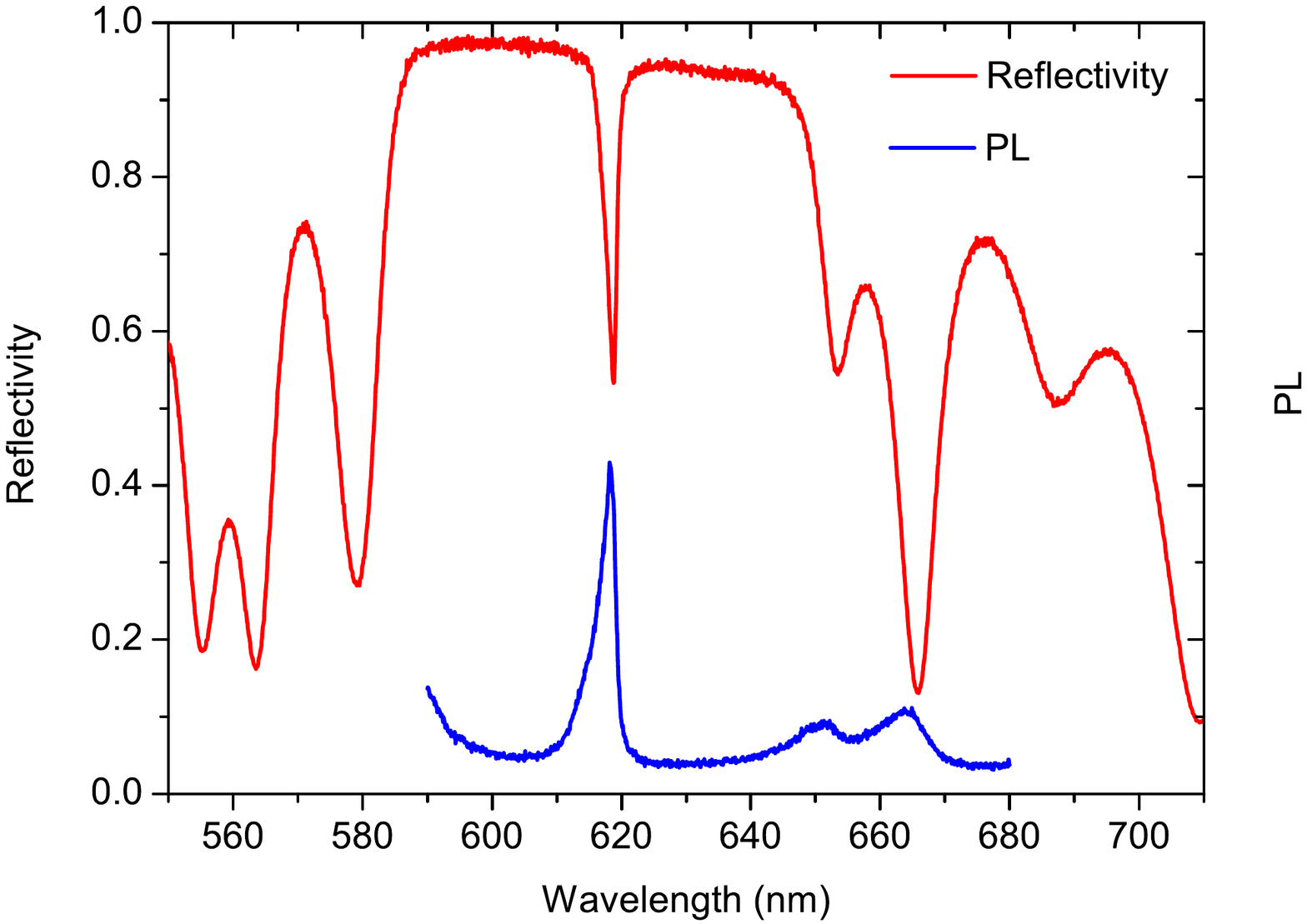}\\
\centering
\caption{(Color online) Photoluminescence and reflectivity spectra measured at $T=2$~K of a planar microcavity based on the mixed Te/Se system. The structure exhibits photoluminescence in the vicinity of the yellow range.}
\label{planar}
\end{figure}

In Fig.~\ref{planar}, are presented the reflectivity and PL spectra of a planar microcavity consisting of 15 and 12 Bragg pairs respectively for the lower and upper DBR. A single CdTe QDs layer has been grown in the $\lambda_0/n$ ZnTe cavity. The cavity mode and photoluminescence reach the vicinity of the yellow range.

    \subsection{Pillar microcavity}

\begin{figure}
  \includegraphics[width=0.6\linewidth]{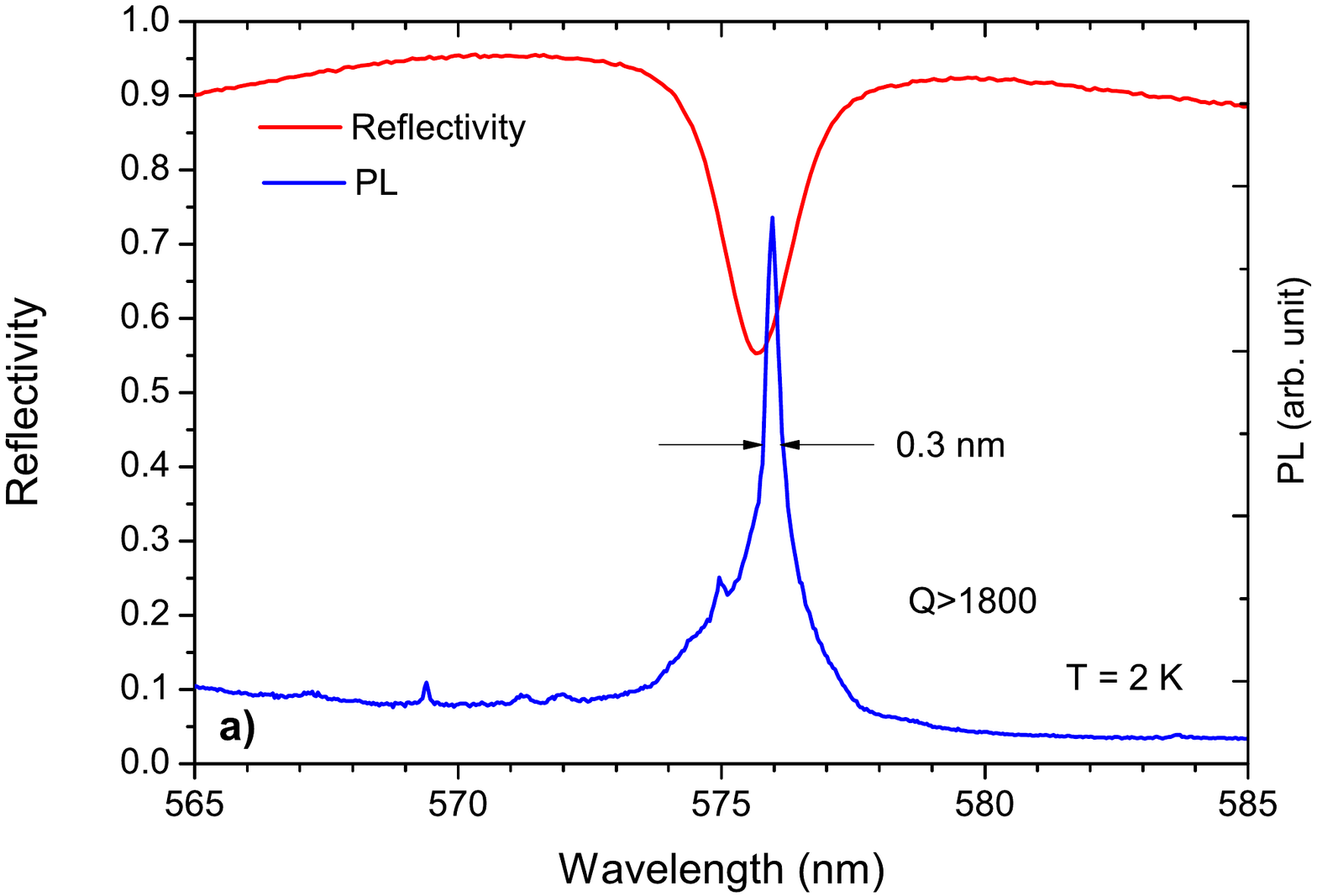}
  \includegraphics[width=0.35\linewidth]{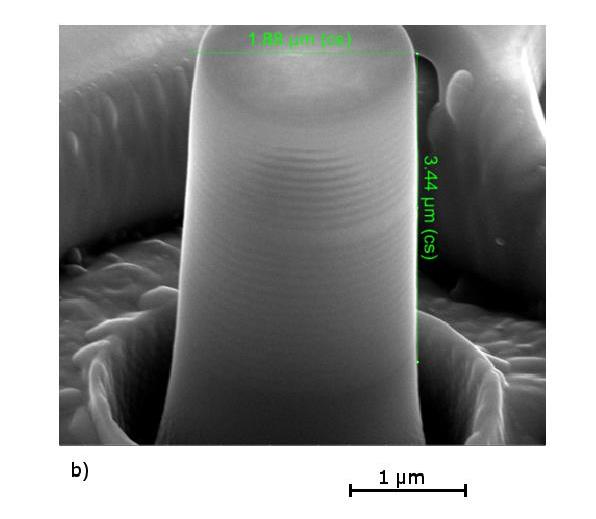}\\
\centering
\caption{(Color online) a) Photoluminescence and reflectivity spectra measured at $T=2$~K of a $2~\mu$m wide pillar etched by FIB. The supplementary lateral confinement of the cavity mode results in a quality factor $Q>1800$ in the yellow range($\lambda=576$~nm) - b) SEM image of the pillar microcavity consisting of  15 and 12 Bragg pairs embedding a $\lambda$ microcavity with a single CdTe/ZnTe QDs layer.}
\label{pillar}
\end{figure}

Micropillars have been etched in the previously presented planar structure, using FIB (focused ion beam) technique (see Fig.~\ref{pillar}). A $2~\mu$m wide pillar exhibit a mode with a quality factor $Q>1800$ in the yellow range ($\lambda=575$~nm). This important result extends the functionalization of such cavities to a range of shorter wavelength and demonstrates the possibility to built VCSELs for yellow optoelectronics.

\section{Conclusion}

In the present work, we report on important new results on microcavities based on selenides and tellurides. We have designed, grown and characterized a microcavity based on selenium compounds only while previous works reported on DBRs only \cite{Peiris_1999} or on structures based on selenides and sulfides \cite{Kruse_2008}. Concerning microcavities based on mixed tellurium and selenium compounds, we have extended the range of application to the yellow/green range (cavity mode obtained at $\lambda=563$~nm) and we characterized a $2~\mu$m pillar cavity exhibiting a quality factor $Q>1800$ at $\lambda=576$~nm, which proves the possibility to design VCSELs for the yellow optoelectronics. Finally, we designed a fully lattice matched microcavity based on tellurium compounds free of Mn in the DBRs, which opens the opportunity to embed dilute magnetic semiconductor layers in a cavity.

\section*{Acknowledgement}

This work was partially supported by the National Center for Research and Development in Poland (project LIDER), the Polish National Science Centre under decisions DEC-2011/02/A/ST3/00131, DEC-2011/01/N/ST3/04536, DEC-2012/05/N/ST3/03209, UMO-2012/05/B/ST7/02155, and by the Foundation for Polish Science. Research was carried out with the use of CePT, CeZaMat and NLTK infrastructures financed by the European Union - the European Regional Development Fund within the Operational Programme "Innovative economy" for 2007-2013.

\end{document}